\documentclass[12pt]{iopart}

\usepackage[english]{babel}

\usepackage[a4paper,top=2cm,bottom=2cm,left=3cm,right=3cm,marginparwidth=1.75cm]{geometry}

\usepackage{longtable}
\usepackage{cite}
\usepackage{amssymb}
\usepackage{graphicx}
\usepackage[colorlinks=true, allcolors=blue]{hyperref}

\expandafter\let\csname equation*\endcsname\relax
\expandafter\let\csname endequation*\endcsname\relax 
\usepackage{amsmath}
\usepackage{gensymb}
\usepackage{glossaries}

\begin{document}

\title[Hydrodynamic Effects in Cryogenic Buffer Gas Cells]{Hydrodynamic Effects in Cryogenic Buffer Gas Cells: Design Insights from Hybrid Simulations}

\author{
Nick Vogeley$^1$, 
Bernd Bauerhenne$^2$, 
Daqing Wang$^1$\footnote{Email: daqing.wang@uni-bonn.de}
}

\vspace{5pt}

\address{$^1$Institute of Applied Physics, University of Bonn, Wegelerstr. 8, 53115 Bonn, Germany}
\address{$^2$Experimentalphysik I, University of Kassel, Heinrich-Plett-Str. 40, 34132, Kassel, Germany}

\vspace{10pt}

\begin{abstract}
Cryogenic buffer gas beam sources have become an essential tool for experiments requiring cold molecular beams with low forward velocities. Although recent experimental advances have led to significant progress in source optimization, numerical studies remain limited due to the challenges posed by the large parameter ranges required to describe both the dense buffer gas and the dilute seed molecules. In this work, we report a numerical evaluation of cryogenic buffer gas beam cells operating in the hydrodynamic extraction regime. While most prior studies focused on box-like or cylindrical cells, we investigated hydrodynamic effects including vortex formation in a spherical cell and assessed whether these could be utilized to enhance the performance in molecule cooling and extraction. To achieve this, we performed steady-state slip-flow simulations for helium buffer gas and employed a direct-simulation Monte Carlo diffusion routine to track particle trajectories. We compared the performance of the source across different buffer gas throughputs and injection angles and identified parameter regimes where vortex formation enhances molecule extraction. From the simulations, we extracted experimental observables, which allow these effects to be verified through velocity or time-of-flight measurements on the molecular beam.

\end{abstract}

\section{Introduction}
Production of cold molecules is essential for a broad spectrum of experimental investigations across multiple disciplines. In particular, cold beams could enable the study of quantum-state-resolved chemical reactions\,\cite{Rios2019, KarmanTomzaRios24}, provide laboratory analogues for astrochemical environments\,\cite{Ziurys2024}, and serve as platforms for quantum information\,\cite{CornishTarbuttHazzard24} and precision spectroscopy\,\cite{DeissWillitschHD24, LangenYe24, DeMilleZelevinsky24}. 
Among various techniques for cold molecular beam generation, cryogenic buffer gas beam (CBGB) sources\,\cite{DoyleCampbell09Book, DoyleHutzler12} have become a workhorse for experiments that require both low forward velocity and low internal temperatures. CBGB source utilizes pre-cooled inert buffer gas, typically helium or neon, to mix with seed molecules in a cryogenic cell. Through collisions, the molecules thermalize with the buffer gas and are subsequently entrained through an orifice, forming a cold molecular beam. An advantage of CBGB over traditional seeded supersonic sources is the forward speed of the beam, which can be lowered to the order of $100\,\mathrm{m\cdot s^{-1}}$, which is particularly advantageous for experiments that require slow beams. The combination of low forward velocities and low internal temperatures has enabled a series of recent experimental advances, including laser cooling and trapping of polyatomic molecules\,\cite{DoyleMitraVilas22, Mitra2020}, spectroscopy of atomic species that lack cycling transitions\,\cite{Hofsass2023,Roeser2024PhysRevA} and organic fluorophores\,\cite{Piskorski2014, Miyamoto2022}.

However, the trade-off is the higher complexity of the experimental setup along with the overhead in maintaining and optimizing the source performance. Despite numerous experimental advances in improving CBGB sources, e.g., in terms of reducing forward velocity\,\cite{White2024} and improving phase space density\,\cite{Wright2023}, there has been only a limited number of numerical investigations\,\cite{singh_optimized_2018,ZeppenfeldGantner20,HutzlerTakahashiShlivko21,Singh2021}. The challenge in simulating buffer gas cells lies in the large mismatch in the parameter regime of the buffer gas and seed molecules in terms of density, pressure and temperature, which precludes the usage of a single solver and necessitates a hybrid approach that combines different simulation methods\,\cite{HutzlerTakahashiShlivko21}. Therefore, a large parameter space of the CBGB design, including cell geometry, buffer gas injection angle, throughput, and orifice design, remains to be explored. Further systematic computational studies are therefore essential to enable the predictive design of CBGB sources and to support their broader applications across a wider range of molecular species.

In this article, we present a hybrid numerical investigation\,\cite{SourceCode} that combines flow dynamics and single-particle tracing to understand and optimize a spherical-shaped CBGB cell. In the flow range of $J\approx20\,\mathrm{sccm}$, we identify the emergence of a hydrodynamical effect that enhances molecular extraction and suppresses contamination of the inner walls of the buffer gas cell. These findings provide a foundation for future experimental validation, e.g., through measurements on the beam velocity and density. We anticipate that this study will contribute to the advancement of numerically guided design strategies for cold molecular beam sources.

\section{Cell Geometry}
\begin{figure}[h]
\includegraphics[width=.45\textwidth,page=2]{figures/250703_CellScheme_angled_bw_paper.pdf}
\includegraphics[width=.45\textwidth,page=1]{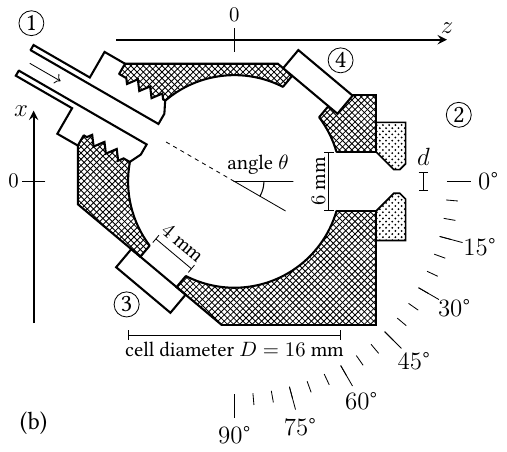}

\caption{Geometry of the simulated buffer gas cell. (a) Three-dimensional rendering of the internal structure of the CBGB cell. The dome-shaped cell has four bores \textcircled{1}, \textcircled{2}, \textcircled{3} and \textcircled{4}, and connects to three external volumes. The volume connected through \textcircled{1} represents the buffer gas inlet and is set to a constant pressure to simulate helium flow into the cell. The volume connected via \textcircled{2} corresponds to the low-pressure region into which the cold beam is extracted. Bore \textcircled{3} represents the molecule injection port. In the simulation, a localized heat load is applied at this position to account for the accompanied heating effects. Opposite to \textcircled{3}, an optical viewport is mounted on bore \textcircled{4}. (b) Cross-sectional view of the buffer gas cell in the $x$-$z$-plane. The exit orifice diameter $d$ and injection angle $\theta$ are varied to investigate their impact on the buffer gas flow dynamics and the molecular diffusion.
}
\label{Fig:Cell}
\end{figure}

The configuration of the buffer gas cell considered in this work is illustrated in Fig.\,\ref{Fig:Cell}. This geometry is conceptualized for easy machinability from a solid copper block using a minimal set of bores. The main body of the cell consists of a cylindrical volume with a diameter of $D=16\,\mathrm{mm}$ and a height of $8\,\mathrm{mm}$. The upper end of the cylinder is enclosed by a hemispherical cap of the same diameter, resulting in an overall structure that resembles an inverted dome. The base of the cylinder is thermally anchored to a cryogenic cold head. Figure\,\ref{Fig:Cell}(a) illustrates a computer-aided design (CAD) model of the buffer gas cell and the affiliated volumes. Throughout this article, we define the symmetry axis of the cylinder as the $y$-axis, while the direction of propagation of the extracted molecular beam is the $z$-axis. Four short, straight cylindrical bores, all lying within the $x$-$z$ plane, are connected to the main cell volume. A cross-sectional view of this plane is shown in Fig.\,\ref{Fig:Cell}(b), where bore \textcircled{1}, with a diameter of $4\,\mathrm{mm}$, serves as the inlet for cold helium buffer gas. Bore \textcircled{2}, having a diameter of $6\,\mathrm{mm}$ and aligned along the $z$-axis, serves as the extraction channel and leads the molecules into the expansion volume. An aperture of diameter $d$ is mounted at the entrance of this bore to define the exit orifice. In the simulations, the angle $\theta$ between the buffer gas injection axis and the $z$-axis is varied in increments of $15\degree$. Additionally, the diameter of the exit aperture $d$ is varied across different simulation runs to permit a comparative analysis of flow dynamics under approximately same flow throughput but at different densities.
Bore \textcircled{3}, also $4\,\mathrm{mm}$ in diameter, represents the injection port for the seed molecules. Located opposite to \textcircled{3} is the bore \textcircled{4} of equal diameter, included to represent a viewport for optical access. The injection angle of seed molecules is fixed at $130\degree$ relative to the $z$-axis.

\section{Simulating and Benchmarking the Helium Flow}
\subsection{Basic settings}
Helium is used as the buffer gas in our simulations. The helium flow is modeled over a range of throughputs from 12 to 87 standard cubic centimeters per minute ($\mathrm{sccm}$), matching the regime of flow rates used in experiments\,\cite{DoylePatterson10}. This flow regime corresponds to Knudsen numbers $\mathrm{Kn} < 0.1$, which permits the use of computational fluid dynamics (CFD) solvers, as also employed in previous work\,\cite{skoff_diffusion_2011,singh_optimized_2018,ZeppenfeldGantner20,HutzlerTakahashiShlivko21}.

We used COMSOL multiphysics with the microfluidics package in the slip flow regime to simulate the buffer gas flow. To do so, the full three-dimensional CAD design of the cell was imported into the simulation environment, where steady-state solutions to the Navier-Stokes equations were solved for helium, modeled as an ideal gas with mass $m_\mathrm{He}=4\,\mathrm{amu}$, ratio of specific heat $\gamma= 5/3$, viscosity $\eta = 1.1078\,\mathrm{\mu Pa\cdot s}$ and heat conductivity $\kappa = 8.636 \,\mathrm{mW \cdot K^{-1} \cdot m^{-1}}$. The transport properties from Ref.\,\cite{HurlyMehl07} were taken considering a reservoir temperature $T_0$ = 4.5\,K. Slip-flow boundary conditions were applied using a total tangential momentum accommodation coefficient $\alpha_t = 0.95$, which is chosen to be higher than the recommended value of 0.9 for technical surfaces at room temperature, and slightly lower than the 0.96 reported for helium interacting with oxygen-coated titanium surfaces\,\cite{JoustenKap9}. The relatively high set value of $\alpha_t = 0.95$ was used to account for the effect that surface roughness causes the particles to scatter more diffusely from walls, bringing the coefficient closer to unity. However, we noted that a few test simulations performed with $\alpha_t$ = 0.90 and 0.99 showed no qualitative impact on the flow behavior. The background pressure of the simulation volume was set to $0.2\,\mathrm{Pa}$, as lower values resulted in numerical instabilities of the CFD solver. While this pressure is relatively high compared to high vacuum conditions typically present in the free-flight region of an experiment, it is not expected to significantly influence the flow pattern within the buffer gas cell.

\subsection{Flow field and temperature}
\begin{figure}
\includegraphics[width=1.\textwidth,page=1]{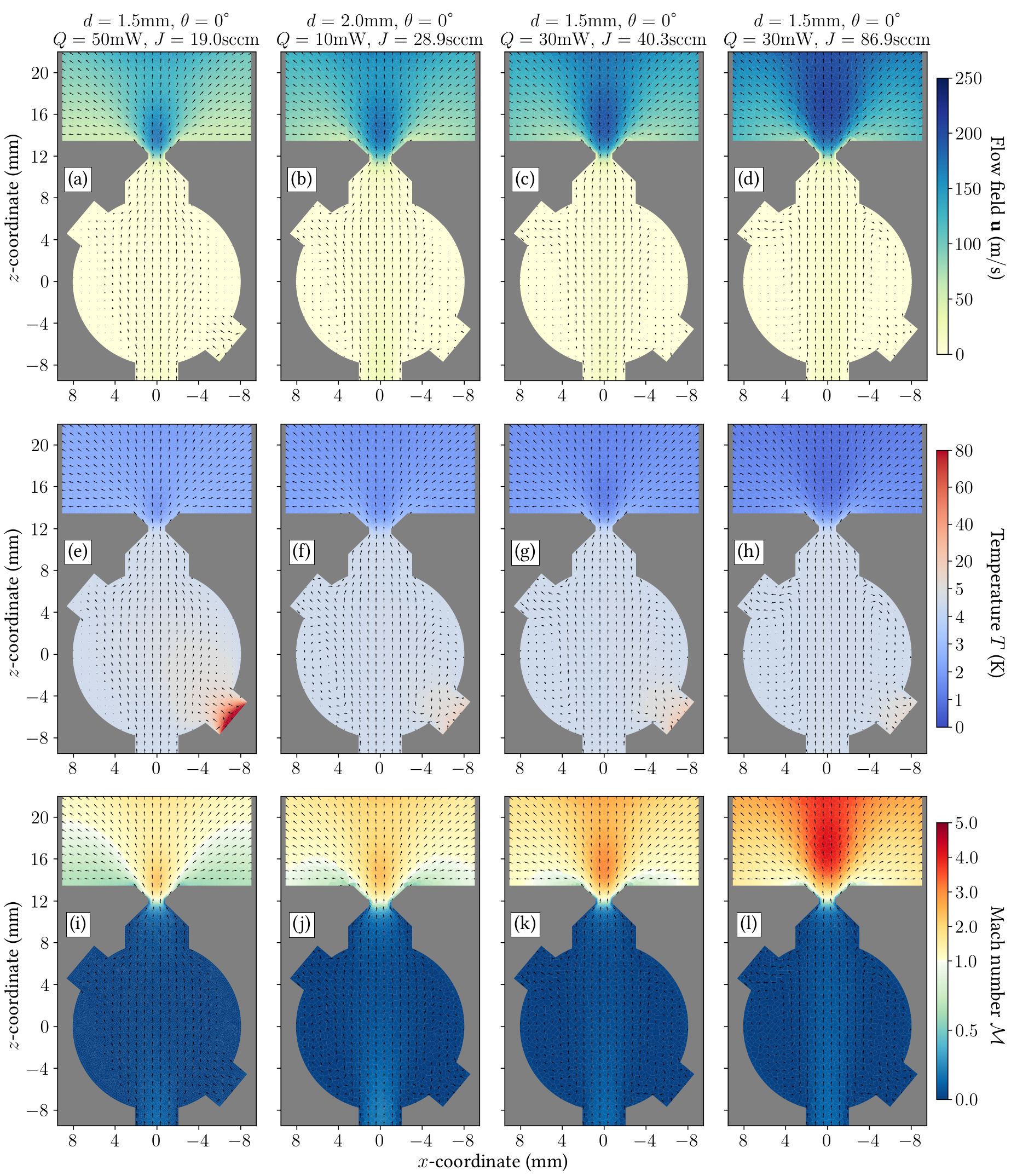}
\caption{Top to bottom: (a)-(d) velocity field $\mathbf{u}$, (e)-(h) temperature $T$ and (i)-(l) Mach number $\mathcal{M}$ distributions on the $y=0$ plane simulated using four different parameter sets, as indicated by the legends on the top rows.}
\label{Fig:Flow_pattern}
\end{figure}

We begin by examining the overall behaviors of the CFD simulation through cross-validation of the resulting distributions of the vector velocity field $\mathbf{u}$, temperature $T$ and flow speed represented by the Mach number $\mathcal{M}$. Each column in Fig.\,\ref{Fig:Flow_pattern} displays the simulation output for $\mathbf{u}$, $T$, and $\mathcal{M}$ (from top to bottom) for the parameter set indicated in the top row of the figure. In total, four different simulations are presented, all performed at a fixed buffer gas injection angle of $\theta=0\degree$, while other parameters including the orifice diameter $d$, external heat load $Q$, and flow throughput $J$ were varied. In each plot, the components of the velocity field $\textbf{u}$ in the $y=0$ plane, $u_x$ and $u_z$, are sampled and displayed as black arrows. The orientation of each arrow indicates the direction of the local flow, while its length is scaled by $\tanh{\sqrt{u_x^2 + u_z^2}}$ to convey a stronger visual contrast of the flow patterns than a simple normalized field. The colored background shows the magnitude of $|\mathbf{u}|$ (upper panel), $T$ (middle panel) and $\mathcal{M}$ (lower panel) on the same $y=0$ plane. Away from the $y=0$ plane, the components $u_x$ and $u_z$ show similar distributions. 

Focusing first on the flow fields depicted in panels (a) to (d), the buffer gas throughput $J$ is gradually increased from 19\,$\mathrm{sccm}$ to 86.9\,$\mathrm{sccm}$ from left to right. As a result, the flow accelerates progressively, as shown by the increasingly darker shading and the lengthening of the arrows. In all plots, the flow within the cell exhibits a gradual acceleration toward the orifice while it remains subsonic. Supersonic expansion is observed at the exit orifice, as indicated by the darker shading in that region. This is accompanied by a noticeable temperature drop in the same region as illustrated in panels (e)–(h) on the middle row, and by increased Mach numbers in panels (i)–(l) on the bottom row. 

The middle panels (e)–(h) display the temperature distributions. The cell walls are set at a constant temperature at $T_0 = 4.5\,\mathrm{K}$, which effectively acts as a thermal reservoir. The flow is expected to behave weakly compressible and quasi-isothermal, which agrees with the observations, where the temperature inside the cell remains nearly uniform at $4.5\,\mathrm{K}$, and only changes notably in the vicinity of the exit orifice. The region with elevated temperature on the bottom right corresponds to the thermal load introduced by the injection of hot molecules. Heat loads of $\{50, 10, 30, 30\}\,\mathrm{mW}$ were applied to the simulations shown in panels (e)–(h), respectively. At the highest applied heat load of $50\,\mathrm{mW}$, the temperature in the injection region reaches up to $80\,\mathrm{K}$, which leads to disturbances in the local flow field. Nevertheless, the impact of this local heating on the flow towards the center of the cell remains negligible.

Panels (i)-(l) show the distribution of Mach numbers. The most prominent feature across all simulations is the onset of supersonic expansion at the orifice. As the buffer gas throughput increases from left to right, the maximal Mach number also increases. A cross examination of the corresponding temperature profiles in panels (e)–(h) reveal the characteristic cooling to temperatures between $2\,\mathrm{K}$ and $0.7\,\mathrm{K}$ associated with adiabatic expansion from the reservoir condition $T_0\approx4.5\,\mathrm{K}$.

The orifice diameter was also varied across the simulations. The two middle columns correspond to simulations with $d=2.0\,\mathrm{mm}$, while the outer columns represent simulations with $d=1.5\,\mathrm{mm}$. A comparison of the Mach number and temperature distributions between the first and second simulations, where similar Mach numbers are achieved, shows that a lower buffer gas throughput of $19.0\,\mathrm{sccm}$ is required for the smaller orifice diameter of $1.5\,\mathrm{mm}$. This observation is consistent with expectations based on conventional supersonic expansion sources.

\subsection{Effect of injection angle and vortex formation}
\begin{figure}
\includegraphics[width=1.\textwidth,page=2]{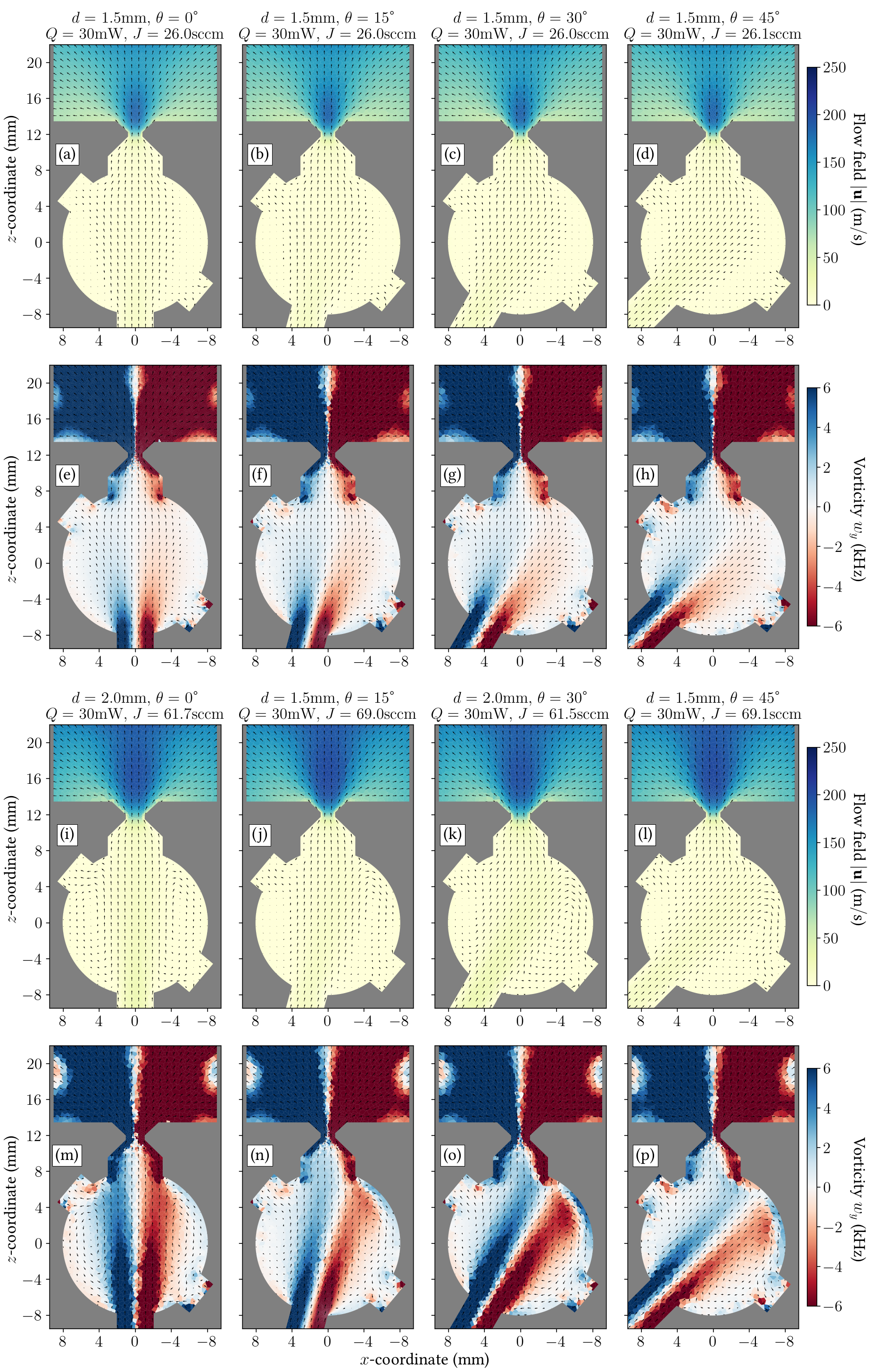}
\caption{Simulated flow fields and vorticity distributions for buffer gas injection angles $\theta = \{0\degree, 15\degree, 30\degree, 45\degree\}$. (a)–(d) show the velocity field for a low-throughput of $26\,\mathrm{sccm}$ and (i)–(l) correspond to simulations with higher throughputs reaching $69\,\mathrm{sccm}$. The respective out-of-plane component $w_y$ of the vorticity field $\mathbf{w}$ are displayed in (e)–(h) and (m)–(p).
}
\label{Fig:Vortex_pattern}
\end{figure}

Having benchmarked key outputs of the CFD simulation, we further investigate the influence of the buffer gas injection angle on the flow field. Panels (a)-(d) of Fig.\,\ref{Fig:Vortex_pattern} display a series of simulations with $d=1.5\,\mathrm{mm}$, $Q=30\,\mathrm{mW}$, $J=26\,\mathrm{sccm}$, and $\theta = \{0\degree, 15\degree, 30\degree, 45\degree\}$. In all cases, the flow efficiently directs helium from the inlet to the orifice, with negligible velocity ($|\mathbf{u}|\approx 0$) near the cell walls and no evidence of backflow. The flow field in the vicinity of the orifice remains largely unaffected by variations in the injection angle. Similarly, no significant changes are observed in the expansion downstream of the orifice.

Panels (i)-(l) depict simulations performed with the same set of injection angles, but with increased buffer gas throughputs reaching $J = 69\,\mathrm{sccm}$. In contrast to flow patterns observed at lower throughputs, distinct vortical structures characterized by regions of backflow near the cell walls, where the axial velocity component satisfies $u_z < 0$, emerge in all configurations. The resulting recirculating flow re-directs the buffer gas inward, effectively refocusing it toward the central region of the cell. As the injection angle increases, the vortical structure becomes more asymmetric and shows a pronounced lobe on the lower right side of the cell, i.e., close to where seed molecules are injected into the cell. Despite these distortions within the cell, the flow pattern in the vicinity of the orifice and in the downstream expansion region remains largely unaffected by variations in the injection angle.

To gain insight into the vortex strength, we extract the vorticity of the flow, defined as $\mathbf{w} = \nabla\times\mathbf{u}$, which characterizes the local rotation of the field. In panels (e)-(h), the $y$-component of vorticity $w_y$ is displayed, which shows strong features in the helium inlet line and changes sign across the center of the pipe. This is consistent with a shear profile in which the velocity magnitude varies transversely, as exemplified in the classical Hagen-Poiseuille flow. A similar sign reversal appears in the expansion region downstream of the orifice, where velocity gradients likewise lead to opposing vorticity components on either side of the central axis. Inside the cell, the vorticity distribution is mostly governed by the injected jet, which inherits characteristics of the helium flow in the pipe. At higher flow throughputs, as shown in panels (m)–(p), more complex structures emerge, which shows as a separation between regions of positive and negative $w_y$ on either side of the centerline, indicating the onset of vortical structures in which the circulation within the two hemispherical regions becomes more pronounced. Of particular interest is the influence of these vortical structures near the injection region of the molecules on the efficiency of molecular extraction, which we will explore in the next section. 

\section{Molecular Tracing through Direct-Simulation Monte Carlo}
The CFD simulations provided an insight into the buffer gas flow characteristics. Due to the low partial pressure of the seed molecules, a fully hydrodynamic treatment of the combined gas mixture is not feasible. To capture the motion of molecules within the background helium flow, we implemented a particle-tracing algorithm in which the center-of-mass motion of molecules is tracked by solving Newton’s equations, with stochastic terms incorporated to model elastic collisions with helium atoms. The helium velocity, density and temperature fields, obtained from the CFD simulations were treated as stationary inputs. This approach enables us to probe the thermalization and transport of molecules within the cell, and quantify the yield of molecule extraction at the orifice. The source code with example simulations is made available\,\cite{SourceCode}.

\subsection{Basic settings}

Specifically, the calculated flow $\mathbf{u}(\mathbf{r})$, number density $n(\mathbf{r})$ and temperature $T(\mathbf{r})$ from the CFD simulation were quantized to a grid of cubic voxels of side lengths $\delta = 500\,\mathrm{\mu m}$. Here, $\mathbf{r}$ denotes the Cartesian coordinates of a given point inside the cell. We tested a few runs with finer meshes of $150\,\mathrm{\mu m}$ but found no systematic deviations besides the increased computational demands.
The values of $\mathbf{u}$, $n$ and $T$ are stored in a look-up table to speed up the calculations. If several data points from the COMSOL mesh fall in the same voxel, their average values are taken. Voxels which did not coincide with mesh points were extrapolated by recursively taking the average of all non-empty nearest neighbors, resulting in a Voronoi-tesselation in Manhattan-metric with soft edges.

Our diffusion model is based on Ref.\,\cite{HutzlerTakahashiShlivko21}. In short, the molecules are modeled as hard spheres of scattering cross section $\sigma = 1.2\times10^{-13}\,\mathrm{cm^2}$ with a mass $m=128\,\mathrm{amu}$, matching the properties of naphthalene as investigated in Ref.\,\cite{DoylePatterson10} and is expected to be sufficiently similar to medium-sized chiral or dye molecules, which could be explored further in subsequent cold beam experiments. The spheres move in straight paths at a velocity $\mathbf{v}$ until an encounter with helium atoms is assigned to the particle. The average collision rate $\Gamma$ is calculated according to
\begin{equation}
    \Gamma = \sigma n(\mathbf{r}) \sqrt{\left|\mathbf{v} - \mathbf{u}(\mathbf{r})\right|^2 + \langle v\rangle_\text{th}^2 }\,,
\end{equation}
which accounts not only for the local thermal velocity of buffer $\langle v\rangle_\text{th}^2 = {8}{k_\mathrm{B}T(\mathbf{r})}/{\pi}{m_\mathrm{He}}$ and instantaneous velocity of the molecule $\mathbf{v}$, but also the local flow velocity $\mathbf{u}(\mathbf{r})$ and number density of buffer gas $n(\mathbf{r})$.

In each simulation step, a molecule moves a distance $\Delta\mathbf{r} = \mathbf{v}\,\Delta t$ with the time step set to $\Delta t = 0.1/\Gamma$. After each time step, a collision was set to occur with a probability of $10\%$. In this way, the interval between two collisions for each individual molecule is allowed to vary drastically while the average is kept constant. As $\mathbf{v}$ changes, so does $\Delta t$ and the mean distance traveled between collisions. When the molecule moves at the same pace and direction as $\mathbf{u}$, it becomes stationary in the co-moving frame of the buffer gas flow and only the thermal motion of the buffer gas induces collisions. In this case, the mean free path $\lambda_\mathrm{m} = v_\mathrm{m}/\Gamma$ recovers the relation $\lambda = 1/(\sigma n \sqrt{ 1 + m/m_\mathrm{He}})$, with the mean relative velocity amplitude $v_\mathrm{m}=\sqrt{8k_BT/\pi m}$\,\cite{Bird, DoyleHutzler12}. We compared two methods for assigning collision partners. First, the velocity of a collision partner was drawn randomly from a Maxwell-Boltzmann distribution. This approach may not account for the impact of the flow field as a moving frame of reference. In a second approach, we generated the three velocity components $\mathbf{v}_\mathrm{th}$ for 10 possible collision partners from a Gaussian distribution with variance $k_\mathrm{B} T(\mathbf{r}) / m$ for each assigned encounter and then drew randomly from this set with relative probabilities $\mathcal{P}\propto |\mathbf{v}_\mathrm{th}+\mathbf{u} - \mathbf{v}|$, to account for the fact that faster particles are more likely to collide.

The back action of the diffusion of molecules on the buffer gas flow was neglected throughout.

\subsection{Molecule trajectory and thermalization}

The injected molecules were initialized at random positions on a circular disc of $4\,\mathrm{mm}$ diameter representing the injection needle, which was modeled with a heat load $Q$ in the COMSOL simulations. The amplitude of the initial velocity vector was sampled from a Maxwell-Boltzmann distribution at $500\,\mathrm{K}$, and the direction was chosen from a $\cos^2(\phi)$ distribution with $\phi$ the polar angle to the normal of the disc. The distribution of a single velocity component $v_i$ with $i={\{x,y,z\}}$ across all initialized molecules is therefore, non-thermal and biased in the direction away from the surface.

After each time step, the position of the molecule is checked. If a trajectory $\mathbf{r}(t)$ moves past the boundary of a voxel, the algorithm checks whether it has moved past the simulation volume. If this is not the case, the local values for $n$, $T$ and $\mathbf{u}$ are updated from the look-up table. If the molecule is found to be out of the simulation volume, the evaluation halts and the particle is considered lost to the wall unless it has reached the exit area.

The molecules were set to propagate for up to $10^9$ collisions and a series of output data containing the time of propagation $t$, position $\mathbf{r}$ and velocity $\mathbf{v}$ was written to a file after every $10^3$ collisions to keep the amount of data manageable. If a wall collision occurred within a threshold number of collisions $\mathrm{K}_\mathrm{th}$, a new particle was initialized to override the previous trajectory. The program was set to run until $10^5$ trajectories with their number of collisions $\mathrm{K}> \mathrm{K}_\mathrm{th}$ were saved.

\begin{figure}

\includegraphics[width=0.48\textwidth,page=1]{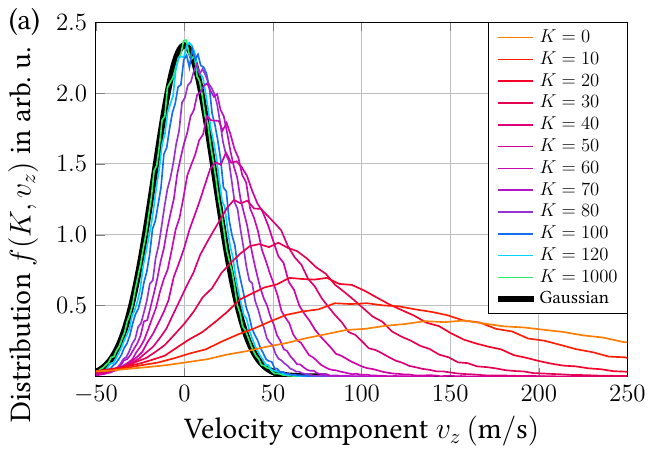}
~\includegraphics[width=0.48\textwidth,page=2]{figures/250709_xyz_velocity_paper.pdf}
\\
\includegraphics[width=0.48\textwidth,page=3]{figures/250709_xyz_velocity_paper.pdf}
~\includegraphics[width=0.48\textwidth]{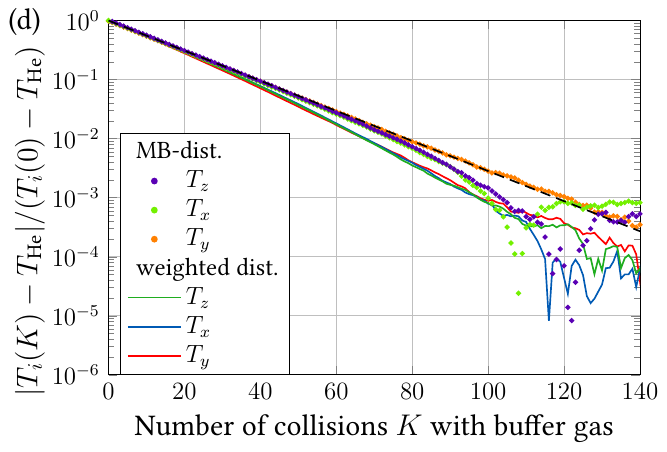}
\caption{(a)-(c) The distribution $f$ of velocity components $v_i$ across all simulated molecules. This data was gathered from simulations with $d = 2.0\,\mathrm{mm}$, $J = 58.8\,\mathrm{sccm}$, helium injection angle of $\theta = 0\degree$ and a cell diameter of 19\,mm. Black lines in the background represent a Gaussian distribution with a temperature of $4.5\,\mathrm{K}$. (d) The relative change of temperature $|T_i(\mathrm{K})-T_\mathrm{He}|/(T_i(0) - T_\mathrm{He})$ plotted as a function of number of collisions experienced.}
\label{Fig:Thermal_velocity}
\end{figure}

To gain insight into the thermalization process, we generated normalized histograms $f(\mathrm{K},v_i) = N(\mathrm{K},v_i)/N(\mathrm{K})$ for each Cartesian velocity component $v_i$ in all $N(\mathrm{K})$ molecules that have experienced $\mathrm{K}$ collisions. In Figs.\,\ref{Fig:Thermal_velocity}(a)-(c), representative histograms for $v_z$, $v_x$, $v_y$ of up to 1000 collisions are displayed. The color code from orange to purple and turquoise represents different values of $\mathrm{K}$ experienced. In $v_z$ and $v_x$, a notable initial offset is present due to the $\cos^2{(\phi)}$ dependence of initial velocity. The maxima of the distributions gradually shift towards lower values and settle at zero with the increasing number of collisions. For $v_y$, the initial distribution $f(\mathrm{K}=0,v_y)$ is centered at zero due to symmetry upon reflection on the $x$-$z$ plane and approaches the familiar bell-curve while staying centered around zero. For all three components, no significant change in statistics was observed after the first $100$ collisions.

By fitting a Gaussian function to each of the histograms, as illustrated by the black lines in Figs.\,\ref{Fig:Thermal_velocity}(a)-(c), we assigned temperatures $T_i(\mathrm{K})$ to the three Cartesian degrees of freedom. For this, we saved the data from the first 200 collisions. The results are presented in Fig.\,\ref{Fig:Thermal_velocity}(d), where the colored lines and dots show the data from the numerical simulations, and the dashed black line follows an analytic model
\begin{equation}
   T_i(K) = T_\mathrm{He} + (T_i(0) - T_\mathrm{He})\exp\left[\dfrac{-2\mathrm{K}m\,m_\mathrm{He}}{(m+m_\mathrm{He})^2}\right]
\end{equation}
provided in Ref.\,\cite{DoyleHutzler12}. With both methods of sampling collision partners, we observe a drop in temperature $|T_i(\mathrm{K}) - T_\mathrm{He}|$ falling within one percent of the initial temperature difference in about 80 collisions. In addition to this global trend, we find that the data of $T_y$ using direct sampling of a Maxwell-Boltzmann distribution recover the analytical formula the best. The reason could be that the buffer gas flow pattern is approximately symmetric upon reflection on the $x$-$z$ plane, while for both two other axes, a mean relative velocity of the helium flow and initial velocity of the molecules is present.

A visible difference of the results using the two sampling methods appears after 40 collisions. The two-step sampling results in faster thermalization. The deviations could hence stem from the fact that the velocity of helium atoms participating in collisions was no longer normally distributed after being randomly sampled twice. With both methods, we observed that within 100 to 140 collisions, the temperature difference between the molecules and the buffer gas reaches the level of $10^{-4}$, hitting the numerical noise floor, which is expected considering our sample size of $10^5$.

\subsection{Extraction efficiency}

\begin{figure}
\includegraphics[width=1.\textwidth,page=1]{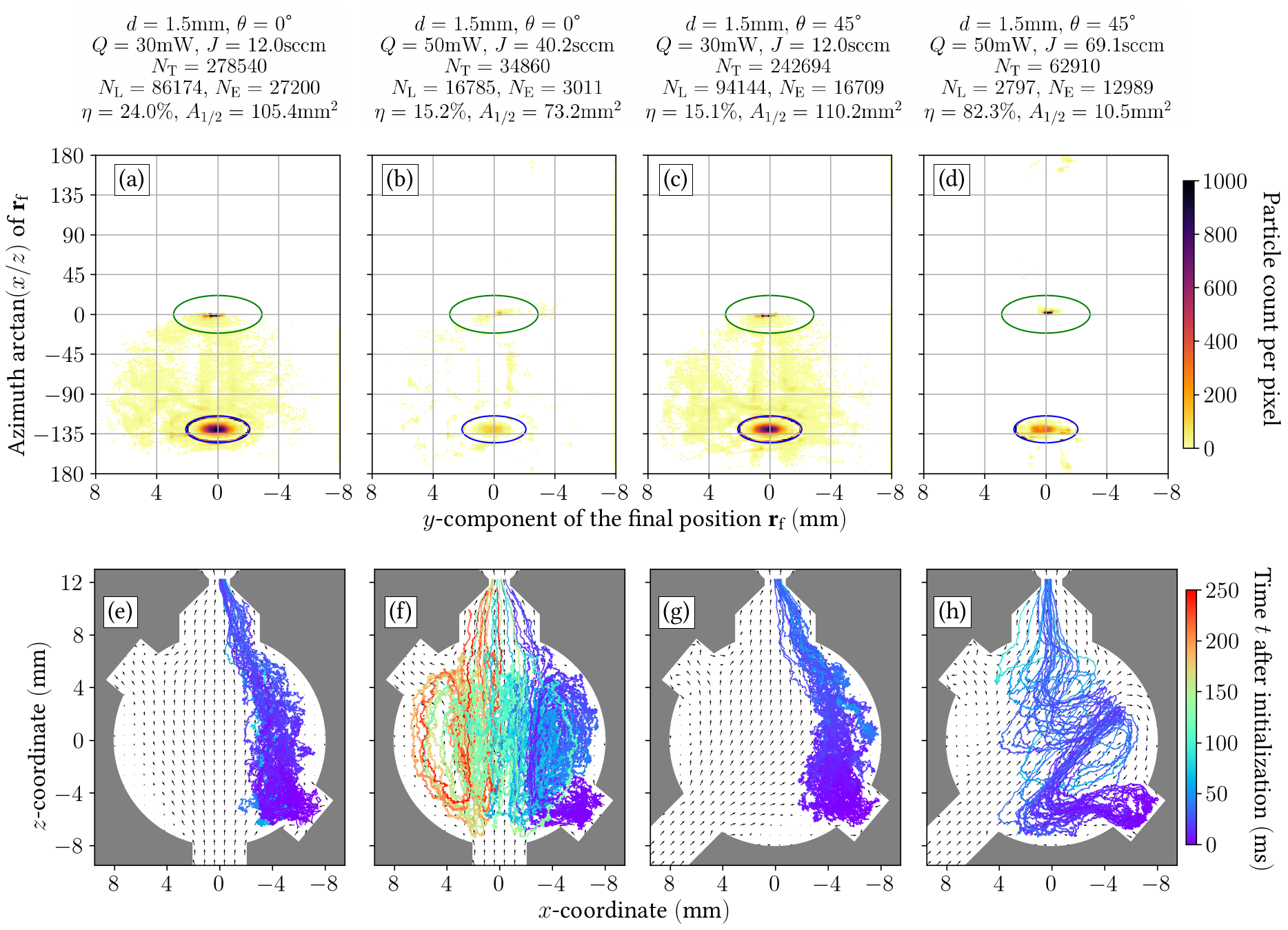}
\caption{(a)-(d) Heat map histograms of last positions $\mathbf{r}_\mathrm{f}$ of molecules. Counts within the green ellipse at $y=0~\mathrm{mm}$, $\arctan{x/z} = 0$ have reached the exit and would participate in the expanding beam. Counts within the blue ellipse $y=0~\mathrm{mm}$, $\arctan{x/z} = -130\degree$ got stuck near their initialization point.
(e)-(h) 25 sample trajectories of molecules moving through the cell. Each curve changes color as a function of time. To keep the displayed data manageable, only positions after every 1000 collisions are shown.}
\label{Fig:Trajectories}
\end{figure}

In the next step, we quantify the extraction efficiency of molecules in cells with different injection angles and operating at various flow throughputs. For each simulation, we evaluated $N_\mathrm{T}\approx 10^5$ trajectories, whose final positions ${\mathbf{r}_\mathrm{f}}$ were extracted. We use a combination of azimuthal angle $\arctan{(x/z)}$ and the $y$-coordinate to uniquely represent $\mathbf{r}_\mathrm{f}$, as the radial distance of almost all final positions equals the radius of the cell. The distributions of $\mathbf{r}_\mathrm{f}$ in four different simulations are shown in the upper row of Fig.\,\ref{Fig:Trajectories}. The molecules can be sorted into three groups. First, molecules that did not leave the source disc. They lie within the blue ellipse centered at $y = 0~\mathrm{mm}$ and $\arctan(x/z) = -130\degree$ in the histogram. This number depended very strongly on the heating power $Q$ and was sensitive to local fluctuations in $\mathbf{u}$. Second, the molecules that reached the exit orifice, corresponding to the ones landed in the area surrounded by the green ellipse. Their number is recorded as $N_\mathrm{E}$. Third, molecules that left the starting region but did not terminate within the exit region are considered losses and contamination of the cell walls. Their number is recorded as $N_\mathrm{L}$.

A first observation of the histograms reveals that a large number of molecules struck the wall within the injection disc shortly after their initialization. This is evident from the relatively strong intensity in the black circles in all plots. The density of molecules deposited on the inner wall is lower compared to that of the inlet and outlet area. Both low-throughput simulations in (a) and (c) feature a broad distribution of molecules between azimuthal angles from $10\degree$ to $-135\degree$ and $y=-4\,$mm to 6\,mm, showing a relatively large amount of wall contamination. Both high-throughput simulations in (b) and (d) show more concentrated distribution in the input and exit areas, hinting at a higher extraction yield. To quantitatively compare different simulations, we define the extraction efficiency
\begin{equation}
    \eta = \dfrac{N_\mathrm{E}}{N_\mathrm{E}+N_\mathrm{L}}
\end{equation}
of only the molecules which were successfully injected into the cell.

Out of surprise, the scenario shown in (d) of with $\theta=45\,\degree$ at $J=69\,\mathrm{sccm}$ gives the highest yield among the four simulations presented. To understand the differences in extraction behavior, we randomly sampled 25 trajectories of extracted molecules from each of the simulations and plotted their projection in the $x$-$z$ plane, as displayed in Figs.\,\ref{Fig:Trajectories}(e)-(h). The change in color along each trajectory represents the time after initialization, as indicated by the color bar. The black arrows in the background indicate the helium flow field.

In the two low-throughput scenarios of (e) and (g), the molecules remained mostly in the half of the cell where they had been injected. Although individual trajectories are rather noisy due to the large number of collisions involved, they follow a general trend towards the exit in a mostly direct path. In stark contrast, the trajectories in the high-throughput regimes displayed in (f) and (h) seem to follow nearly helical paths. While this may seem less efficient at first, the configuration $\theta = 45\degree$ at $J=69\,\mathrm{sccm}$ outperforms all low-throughput simulations, where $\eta$ reaches around 80\%, which means eight of ten molecules that escaped the source disc arrive at the exit. In this scenario, shown in (h), the molecules are first drawn away from the source area by the strong vortex field toward the entrance of helium gas and then pushed straight through the center of the cell towards the orifice.

\begin{figure}
\includegraphics[width=1\textwidth,page=1]{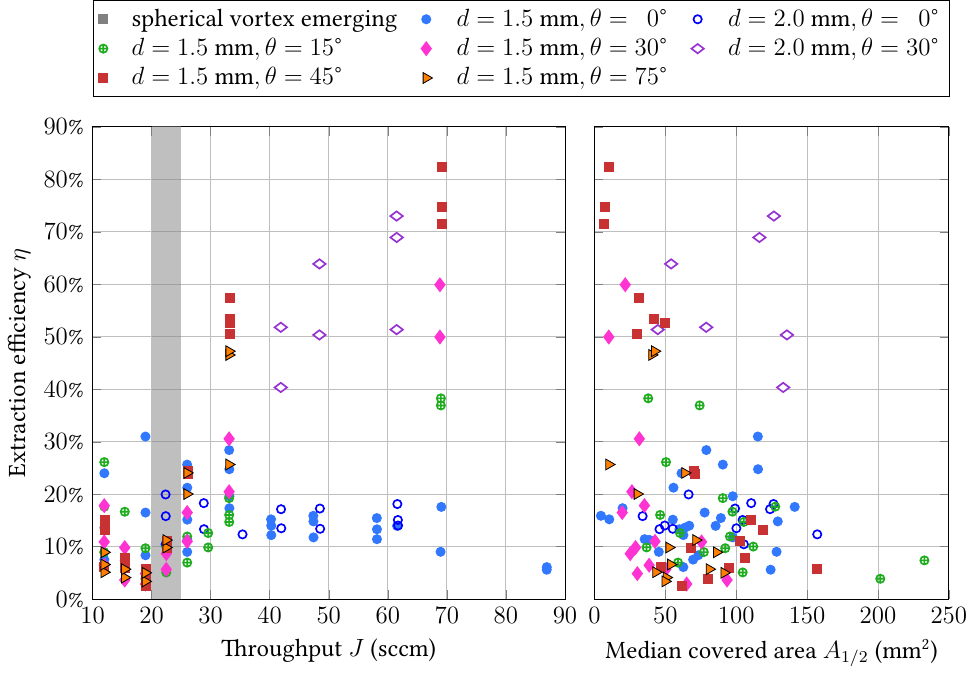}
\caption{Extraction efficiency $\eta$ plotted over throughput $J$ and median covered area $A_{1/2}$. The gray vertical bar illustrate the flow range where vortices in the buffer gas flow emerge. For $\theta = 0\degree$, efficiency goes down with $J$ in an approximately linear fashion. For $\theta > 30\degree$, a change in trend is observed after the onset of vortex formation around $J \approx 25\,\mathrm{sccm}$ with the efficiency increasing and $A_{1/2}$ decreasing for higher throughput.}
\label{Fig:Yield_Throughput}
\end{figure}

A summary of the extraction efficiencies obtained from a total of 115 simulations is presented in the left panel of Fig.\,\ref{Fig:Yield_Throughput} and detailed statistics on these simulations is listed in \ref{Appendix}. In the low-throughput regime ($J < 30\,\mathrm{sccm}$), all geometries show low extraction efficiencies, mostly remaining below 25\%. As $J$ increases, configurations with inclined helium injection show a notable rise in extraction efficiency. In particular, three separate simulations with $\theta=45\,\degree$ at $J =69\,\mathrm{sccm}$ resulted in extraction yield all above 70\%. Parallel to the extraction yield, we analyzed the equivalent coated median area $A_{1/2}$ of the pixels that make up $N_\mathrm{L}/2$ of the molecules that landed on the inner surface of the cell. A smaller $A_{1/2}$ implies a reduced spread across the cell walls. The plot in the right panel of Fig.\,\ref{Fig:Yield_Throughput} confirms that the configuration of $\theta=45\,\degree$ in the high-throughput regime also shows minimal wall contamination.

\begin{figure}
\includegraphics[width=1.\textwidth,page=1]{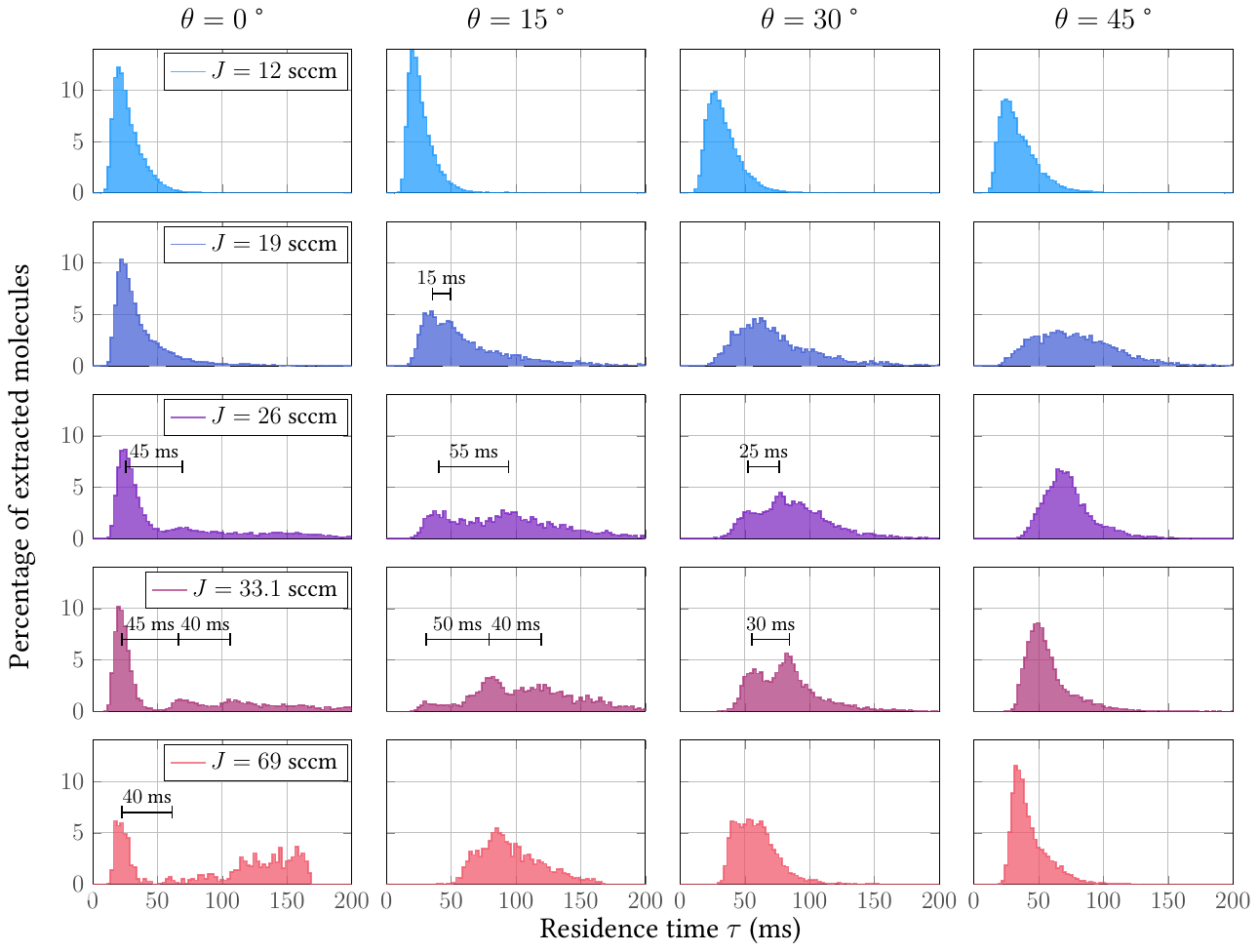}
\caption{Histogram of residence times $\tau$ for different geometries $\theta$ with fixed $d=1.5~\mathrm{mm}$ at different throughput $J$. With the emergence of the spherical vortex, several maxima appear beside the main peak which stays relatively consistent around $\tau \approx 25\,\mathrm{ms}$ but changes in intensity. The new maxima appear regularly spaced, hinting at molecules traveling along the vortex for one or two additional full loops before exiting. At the steeper injection angle $\theta=45{\degree}$, this effect is reduced, possibly due to the wide side lobe of the vortex facing the injection side. The emergence of two or three beam intensity maxima after pulsed injection of molecules could be verified experimentally by time-of-flight measurements.}
\label{Fig:Extraction_Time}
\end{figure}

\subsection{Extraction time}
In the final step, we analyze the residence time $\tau$ in the cell for all successfully extracted molecules. Figure\,\ref{Fig:Extraction_Time} presents the normalized histograms of residence time for simulations performed at four different injection angles, each evaluated at four different flow throughputs. In the low-throughput regime ($J = 12\,\mathrm{sccm}$), different cell geometries result in similar extraction-time distributions, with the majority of molecules exiting the cell within 10\,ms to 50\,ms. As the flow rate increases, the geometry with $\theta = 0\degree$ exhibits two additional peaks in the residence time distribution. The emergence of these peaks coincides with the flow regime of vortex formation in the buffer gas flow. A further observation of that the three maxima are close to evenly spaced could be the result of an integer number of revolutions in one of the vortex lobes before being extracted. This behavior is corroborated by the molecular trajectories shown in Fig.\,\ref{Fig:Trajectories}(f). The extraction time for the majority of molecules remains peaked at approximately 20\,ms, with the peak position shifting slightly toward shorter times as the throughput increases, which is consistent with the corresponding rise in flow velocity.

For the scenarios of $\theta=15\degree$ and $30\degree$, similar multi-peak distributions are observed. However, the maxima of the distributions switch to peaks at long time scales in the simulations with $J=26\,\mathrm{sccm}$ and 31\,$\mathrm{sccm}$. As the flow rate increases further, the peaks merge and a broad distribution takes over with the peak extraction time shifted to around 50\,$\mathrm{ms}$ to 80\,$\mathrm{ms}$. This signifies that vortices in the buffer gas flow lead to delays in extraction and a larger uncertainty in the residence time.

For geometry with $\theta = 45\degree$, no multi-peaked residence time distribution is observed. At throughputs above 19\,$\mathrm{sccm}$, increasing $J$ results in a progressive narrowing of the distribution and a shift of the peak residence time toward approximately 30\,$\mathrm{ms}$, i.e., a focusing effect that both narrows down the distribution and allows fast extraction. This observation is in accordance with the increased extraction yield discussed in the previous section.

\section{Conclusion}
In summary, we have presented the numerical investigation of hydrodynamic effects in a cryogenic buffer gas beam source. A hybrid approach combining CFD simulations and particle tracing provided detailed insights into the buffer gas flow pattern, as well as the thermalization and extraction behaviors of the seed molecules. The results show that key parameters such as the injection angle of the buffer gas and flow rate have subtle effects on the extraction efficiency. The geometry with a helium injection angle at $45\degree$ in the high-throughput regime shows unexpectedly advantageous features, including high extraction efficiency and narrow extraction time distributions. These results could be tested through velocity and time-of-flight measurements downstream of the cell in future experimental studies. While the flow rates considered in this work are at the upper limit of those reported in recent experimental setups\,\cite{PattersonDoyle2007,DoylePatterson10}, the complete throughput regime should become accessible with increased pumping capacity. More generally, we hope that the numerical tools presented in this work\,\cite{SourceCode} can assist future optimization work of buffer gas sources, including, for example, the design of more sophisticated sources involving multi-stages\,\cite{White2024,Singh2021} to reduce forward velocities and further lower the internal temperatures.

\section*{Acknowledgement}
This work was funded by the Deutsche Forschungsgemeinschaft (DFG, German Research Foundation) — Project No. 328961117 — SFB 1319 ELCH (Extreme light for sensing and driving molecular chirality) and under Germany’s Excellence Strategy — Cluster of Excellence Matter and Light for Quantum Computing (ML4Q) EXC 2004/1 - 390534769.

\vspace{1cm}
\section*{References}
\bibliographystyle{vancouver}
\bibliography{sources}

\appendix
\section{Parameters and statistics of all particle trace simulations}
\label{Appendix}
{\footnotesize
\begin{longtable}{r||r|r|r||r|r|r|r||r|r|r}
\caption{Input parameters and key output results for all the particle tracing simulations. Geometric parameters $\theta$, $d$, and simulation parameters $Q$, $J$ and mean buffer gas density in the cell $\langle n\rangle$ listed with corresponding particle tracing parameters $N_\mathrm{T}$, $N_\mathrm{L}$, $N_\mathrm{E}$ and calculated mean extraction time $\langle\tau\rangle$, median covered area $A_{1/2}$ and extraction efficiency $\eta$. All the 115 simulations are available on Zenodo\,\cite{SourceCode}.\label{Tab:Yields}}
\\
    $\theta$&	$d$&	$Q$&	$J$&    $\langle n\rangle$
	&
	$N_\mathrm{T}$&   $N_\mathrm{L}$&	$N_\mathrm{E}$
	&
	$\langle\tau\rangle$&	$A_{1/2}$&	$\eta$
	\\\hline
	&
	($\mathrm{mm}$)
	&
	($\mathrm{mW}$)
	&
	($\mathrm{sccm}$)
	&
	($\mathrm{cm}^{-3}$)
	&
	
	&
	
	&
	
	&
	($\mathrm{ms}$)
	&
	($\mathrm{mm}^2$)
	&
	$\%$
	\\\hline\hline
$0$°&	$1.5$&	$5$&	$12.00$&	$6.25\cdot10^{16}$&	$1390256$&	$97706$&	$7963$& 	$78$&	$70$&	$7.5$\\
$0$°&	$1.5$&	$10$&	$12.01$&	$6.22\cdot10^{16}$&	$446589$&	$107955$&	$17029$&	$135$&	$64$&	$13.6$\\
$0$°&	$1.5$&	$30$&	$12.00$&	$5.99\cdot10^{16}$&	$278540$&	$86174$&	$27200$&	$123$&	$61$&	$24$\\
$0$°&	$1.5$&	$10$&	$18.97$&	$9.45\cdot10^{16}$&	$503491$&	$62484$&	$5705$& 	$112$&	$73$&	$8.4$\\
$0$°&	$1.5$&	$30$&	$18.97$&	$9.26\cdot10^{16}$&	$186096$&	$82149$&	$16237$&	$82$&	$78$&	$16.5$\\
$0$°&	$1.5$&	$50$&	$18.96$&	$8.96\cdot10^{16}$&	$212883$&	$72466$&	$32534$&	$72$&	$115$&	$31$\\
$0$°&	$1.5$&	$5$&	$26.05$&	$1.27\cdot10^{17}$&	$674900$&	$8136$& 	$1036$& 	$150$&	$39$&	$11.3$\\
$0$°&	$1.5$&	$10$&	$26.03$&	$1.27\cdot10^{17}$&	$410594$&	$24186$&	$2395$& 	$135$&	$45$&	$9.0$\\
$0$°&	$1.5$&	$10$&	$26.05$&	$1.27\cdot10^{17}$&	$423709$&	$23367$&	$4171$& 	$127$&	$55$&	$15.1$\\
$0$°&	$1.5$&	$30$&	$26.05$&	$1.25\cdot10^{17}$&	$97651$&	$41224$&	$11115$&	$98$&	$56$&	$21$\\
$0$°&	$1.5$&	$50$&	$26.02$&	$1.23\cdot10^{17}$&	$100507$&	$43196$&	$14890$&	$85$&	$91$&	$26$\\
$0$°&	$1.5$&	$5$&	$33.18$&	$1.59\cdot10^{17}$&	$519723$&	$3472$& 	$728$&  	$138$&	$20$&	$17.3$\\
$0$°&	$1.5$&	$10$&	$33.18$&	$1.59\cdot10^{17}$&	$453490$&	$10256$&	$2501$& 	$134$&	$98$&	$19.6$\\
$0$°&	$1.5$&	$30$&	$33.17$&	$1.58\cdot10^{17}$&	$92320$&	$27264$&	$8970$& 	$133$&	$115$&	$25$\\
$0$°&	$1.5$&	$50$&	$33.14$&	$1.56\cdot10^{17}$&	$67058$&	$27680$&	$10983$&	$120$&	$79$&	$28$\\
$0$°&	$1.5$&	$10$&	$40.27$&	$1.91\cdot10^{17}$&	$276370$&	$4963$& 	$690$&  	$113$&	$63$&	$12.2$\\
$0$°&	$1.5$&	$30$&	$40.26$&	$1.90\cdot10^{17}$&	$57742$&	$15449$&	$2514$& 	$97$&	$67$&	$14.0$\\
$0$°&	$1.5$&	$50$&	$40.23$&	$1.88\cdot10^{17}$&	$34860$&	$16785$&	$3011$& 	$135$&	$10.4$&	$15.2$\\
$0$°&	$1.5$&	$10$&	$47.42$&	$2.23\cdot10^{17}$&	$252658$&	$2451$& 	$327$&  	$95$&	$98$&	$11.8$\\
$0$°&	$1.5$&	$30$&	$47.42$&	$2.22\cdot10^{17}$&	$63019$&	$11541$&	$2007$& 	$111$&	$129$&	$14.8$\\
$0$°&	$1.5$&	$50$&	$47.39$&	$2.21\cdot10^{17}$&	$30832$&	$13170$&	$2487$& 	$111$&	$4.6$&	$15.9$\\
$0$°&	$1.5$&	$10$&	$58.17$&	$2.71\cdot10^{17}$&	$205100$&	$1038$& 	$134$&  	$110$&	$35$&	$11.4$\\
$0$°&	$1.5$&	$30$&	$58.16$&	$2.70\cdot10^{17}$&	$83936$&	$8129$& 	$1249$& 	$45$&	$60$&	$13.3$\\
$0$°&	$1.5$&	$50$&	$58.14$&	$2.69\cdot10^{17}$&	$30503$&	$9861$& 	$1802$& 	$41$&	$89$&	$15.5$\\
$0$°&	$1.5$&	$10$&	$68.94$&	$3.18\cdot10^{17}$&	$170229$&	$514$&  	$51$&   	$28$&	$128$&	$9.0$\\
$0$°&	$1.5$&	$30$&	$61.57$&	$3.19\cdot10^{17}$&	$136222$&	$6712$& 	$1089$& 	$55$&	$85$&	$14.0$\\
$0$°&	$1.5$&	$50$&	$69.07$&	$3.18\cdot10^{17}$&	$40831$&	$9682$& 	$2065$& 	$38$&	$141$&	$17.6$\\
$0$°&	$1.5$&	$10$&	$86.92$&	$3.98\cdot10^{17}$&	$125713$&	$215$&  	$14$&   	$89$&	$63$&	$6.1$\\
$0$°&	$1.5$&	$30$&	$86.92$&	$3.98\cdot10^{17}$&	$123289$&	$2104$& 	$125$&  	$64$&	$124$&	$5.6$\\
$0$°&	$2.0$&	$10$&	$22.40$&	$6.10\cdot10^{16}$&	$301096$&	$79187$&	$14878$&	$45$&	$34$&	$15.8$\\
$0$°&	$2.0$&	$30$&	$22.37$&	$5.88\cdot10^{16}$&	$248630$&	$82246$&	$20493$&	$39$&	$66$&	$19.9$\\
$0$°&	$2.0$&	$5$&	$22.39$&	$6.12\cdot10^{16}$&	$641844$&	$54046$&	$6309$& 	$29$&	$105$&	$10.5$\\
$0$°&	$2.0$&	$10$&	$28.87$&	$7.70\cdot10^{16}$&	$240742$&	$48489$&	$7460$& 	$52$&	$46$&	$13.3$\\
$0$°&	$2.0$&	$30$&	$28.84$&	$7.52\cdot10^{16}$&	$128037$&	$51533$&	$11540$&	$38$&	$111$&	$18.3$\\
$0$°&	$2.0$&	$10$&	$35.40$&	$9.29\cdot10^{16}$&	$186236$&	$28962$&	$4080$& 	$33$&	$157$&	$12.3$\\
$0$°&	$2.0$&	$10$&	$41.96$&	$1.09\cdot10^{17}$&	$205171$&	$20255$&	$3164$& 	$33$&	$100$&	$13.5$\\
$0$°&	$2.0$&	$30$&	$41.92$&	$1.07\cdot10^{17}$&	$60302$&	$27078$&	$5602$& 	$28$&	$124$&	$17.1$\\
$0$°&	$2.0$&	$10$&	$48.54$&	$1.24\cdot10^{17}$&	$203310$&	$13918$&	$2155$& 	$36$&	$55$&	$13.4$\\
$0$°&	$2.0$&	$30$&	$48.50$&	$1.23\cdot10^{17}$&	$48897$&	$21322$&	$4450$& 	$70$&	$99$&	$17.3$\\
$0$°&	$2.0$&	$10$&	$61.74$&	$1.56\cdot10^{17}$&	$196596$&	$6440$& 	$1051$& 	$73$&	$50$&	$14.0$\\
$0$°&	$2.0$&	$30$&	$61.71$&	$1.55\cdot10^{17}$&	$34084$&	$13742$&	$2442$& 	$81$&	$105$&	$15.1$\\
$0$°&	$2.0$&	$50$&	$61.62$&	$1.53\cdot10^{17}$&	$29636$&	$14569$&	$3225$& 	$65$&	$126$&	$18.1$\\
$15$°&	$1.5$&	$5$&	$11.98$&	$6.25\cdot10^{16}$&	$849538$&	$59968$&	$5903$& 	$103$&	$77$&	$9.0$\\
$15$°&	$1.5$&	$10$&	$11.99$&	$6.22\cdot10^{16}$&	$378291$&	$93373$&	$19973$&	$88$&	$128$&	$17.6$\\
$15$°&	$1.5$&	$30$&	$11.99$&	$6.00\cdot10^{16}$&	$210648$&	$78020$&	$27574$&	$129$&	$50$&	$26$\\
$15$°&	$1.5$&	$10$&	$15.46$&	$7.84\cdot10^{16}$&	$307772$&	$65179$&	$5199$& 	$96$&	$233$&	$7.4$\\
$15$°&	$1.5$&	$30$&	$15.46$&	$7.64\cdot10^{16}$&	$151923$&	$70967$&	$14186$&	$141$&	$97$&	$16.7$\\
$15$°&	$1.5$&	$10$&	$18.96$&	$9.45\cdot10^{16}$&	$292436$&	$38122$&	$1542$& 	$111$&	$202$&	$3.9$\\
$15$°&	$1.5$&	$30$&	$18.97$&	$9.28\cdot10^{16}$&	$84827$&	$45558$&	$4896$& 	$130$&	$92$&	$9.7$\\
$15$°&	$1.5$&	$10$&	$22.49$&	$1.11\cdot10^{17}$&	$282795$&	$21116$&	$1133$& 	$110$&	$105$&	$5.1$\\
$15$°&	$1.5$&	$30$&	$22.49$&	$1.09\cdot10^{17}$&	$51531$&	$27794$&	$3095$& 	$97$&	$112$&	$10.0$\\
$15$°&	$1.5$&	$10$&	$26.03$&	$1.27\cdot10^{17}$&	$265554$&	$12152$&	$910$&  	$97$&	$59$&	$7.0$\\
$15$°&	$1.5$&	$30$&	$26.03$&	$1.25\cdot10^{17}$&	$39484$&	$20677$&	$2802$& 	$93$&	$96$&	$11.9$\\
$15$°&	$1.5$&	$10$&	$29.59$&	$1.43\cdot10^{17}$&	$269103$&	$7315$& 	$800$&  	$48$&	$37$&	$9.9$\\
$15$°&	$1.5$&	$30$&	$29.59$&	$1.42\cdot10^{17}$&	$35444$&	$17249$&	$2486$& 	$44$&	$60$&	$12.6$\\
$15$°&	$1.5$&	$10$&	$33.15$&	$1.59\cdot10^{17}$&	$430029$&	$6985$& 	$1205$& 	$35$&	$105$&	$14.7$\\
$15$°&	$1.5$&	$30$&	$33.15$&	$1.58\cdot10^{17}$&	$59091$&	$22924$&	$4381$& 	$62$&	$46$&	$16.0$\\
$15$°&	$1.5$&	$50$&	$33.13$&	$1.56\cdot10^{17}$&	$44586$&	$22699$&	$5412$& 	$46$&	$91$&	$19.3$\\
$15$°&	$1.5$&	$30$&	$69.00$&	$3.19\cdot10^{17}$&	$136221$&	$4213$& 	$2611$& 	$88$&	$38$&	$38$\\
$15$°&	$1.5$&	$50$&	$68.99$&	$3.18\cdot10^{17}$&	$39740$&	$7693$& 	$4501$& 	$75$&	$74$&	$37$\\
$30$°&	$1.5$&	$5$&	$11.98$&	$6.25\cdot10^{16}$&	$837304$&	$76668$&	$5301$& 	$98$&	$39$&	$6.5$\\
$30$°&	$1.5$&	$10$&	$11.99$&	$6.23\cdot10^{16}$&	$407908$&	$110141$&	$13485$&	$89$&	$76$&	$10.9$\\
$30$°&	$1.5$&	$30$&	$11.99$&	$6.03\cdot10^{16}$&	$282836$&	$94635$&	$20538$&	$98$&	$35$&	$17.8$\\
$30$°&	$1.5$&	$10$&	$15.45$&	$7.85\cdot10^{16}$&	$380417$&	$81867$&	$3103$& 	$88$&	$94$&	$3.7$\\
$30$°&	$1.5$&	$30$&	$15.46$&	$7.68\cdot10^{16}$&	$196777$&	$87819$&	$9579$& 	$93$&	$29$&	$9.8$\\
$30$°&	$1.5$&	$10$&	$18.96$&	$9.46\cdot10^{16}$&	$295186$&	$48391$&	$1440$& 	$83$&	$65$&	$2.9$\\
$30$°&	$1.5$&	$30$&	$18.96$&	$9.32\cdot10^{16}$&	$120554$&	$61590$&	$3146$& 	$49$&	$30$&	$4.8$\\
$30$°&	$1.5$&	$10$&	$22.49$&	$1.11\cdot10^{17}$&	$253987$&	$29212$&	$1757$& 	$42$&	$50$&	$5.7$\\
$30$°&	$1.5$&	$30$&	$22.49$&	$1.10\cdot10^{17}$&	$71110$&	$35967$&	$3408$& 	$47$&	$25$&	$8.7$\\
$30$°&	$1.5$&	$10$&	$26.03$&	$1.27\cdot10^{17}$&	$229750$&	$18337$&	$2268$& 	$39$&	$43$&	$11.0$\\
$30$°&	$1.5$&	$30$&	$26.03$&	$1.26\cdot10^{17}$&	$51833$&	$23680$&	$4678$& 	$43$&	$20$&	$16.5$\\
$30$°&	$1.5$&	$10$&	$33.13$&	$1.59\cdot10^{17}$&	$254923$&	$7786$& 	$2004$& 	$36$&	$26$&	$20$\\
$30$°&	$1.5$&	$30$&	$33.13$&	$1.58\cdot10^{17}$&	$42481$&	$14978$&	$6588$& 	$32$&	$32$&	$31$\\
$30$°&	$1.5$&	$30$&	$68.83$&	$3.19\cdot10^{17}$&	$99461$&	$4251$& 	$4241$& 	$60$&	$10.1$&	$50$\\
$30$°&	$1.5$&	$50$&	$68.83$&	$3.18\cdot10^{17}$&	$40071$&	$5603$& 	$8369$& 	$56$&	$22$&	$60$\\
$30$°&	$2.0$&	$10$&	$41.88$&	$1.09\cdot10^{17}$&	$539833$&	$40169$&	$27154$&	$57$&	$133$&	$40$\\
$30$°&	$2.0$&	$30$&	$41.89$&	$1.08\cdot10^{17}$&	$173132$&	$48571$&	$52188$&	$44$&	$79$&	$52$\\
$30$°&	$2.0$&	$10$&	$48.42$&	$1.25\cdot10^{17}$&	$511826$&	$23439$&	$23761$&	$38$&	$136$&	$50$\\
$30$°&	$2.0$&	$30$&	$48.43$&	$1.24\cdot10^{17}$&	$148931$&	$31506$&	$55685$&	$85$&	$54$&	$64$\\
$30$°&	$2.0$&	$10$&	$61.50$&	$1.56\cdot10^{17}$&	$475344$&	$11488$&	$12128$&	$86$&	$45$&	$51$\\
$30$°&	$2.0$&	$30$&	$61.51$&	$1.56\cdot10^{17}$&	$158521$&	$23399$&	$51857$&	$75$&	$116$&	$69$\\
$30$°&	$2.0$&	$50$&	$61.49$&	$1.54\cdot10^{17}$&	$133082$&	$23047$&	$62265$&	$67$&	$127$&	$73$\\
$45$°&	$1.5$&	$0$&	$12.02$&	$6.28\cdot10^{16}$&	$1572851$&	$10250$&	$659$&  	$53$&	$47$&	$6.0$\\
$45$°&	$1.5$&	$10$&	$12.04$&	$6.25\cdot10^{16}$&	$340580$&	$95893$&	$14505$&	$42$&	$119$&	$13.1$\\
$45$°&	$1.5$&	$30$&	$12.04$&	$6.04\cdot10^{16}$&	$242694$&	$94144$&	$16709$&	$36$&	$110$&	$15.1$\\
$45$°&	$1.5$&	$10$&	$15.52$&	$7.86\cdot10^{16}$&	$423284$&	$75130$&	$4819$& 	$59$&	$95$&	$6.0$\\
$45$°&	$1.5$&	$30$&	$15.52$&	$7.68\cdot10^{16}$&	$204584$&	$100972$&	$8619$& 	$46$&	$106$&	$7.9$\\
$45$°&	$1.5$&	$0$&	$19.02$&	$9.49\cdot10^{16}$&	$878386$&	$5088$& 	$126$&  	$93$&	$62$&	$2.4$\\
$45$°&	$1.5$&	$10$&	$19.03$&	$9.48\cdot10^{16}$&	$425021$&	$53683$&	$2145$& 	$87$&	$80$&	$3.8$\\
$45$°&	$1.5$&	$30$&	$19.04$&	$9.33\cdot10^{16}$&	$119523$&	$66396$&	$4031$& 	$79$&	$157$&	$5.7$\\
$45$°&	$1.5$&	$10$&	$22.57$&	$1.11\cdot10^{17}$&	$342664$&	$32101$&	$3464$& 	$92$&	$68$&	$9.7$\\
$45$°&	$1.5$&	$30$&	$22.58$&	$1.10\cdot10^{17}$&	$80006$&	$41132$&	$5109$& 	$88$&	$103$&	$11.1$\\
$45$°&	$1.5$&	$10$&	$26.13$&	$1.27\cdot10^{17}$&	$268700$&	$18011$&	$5847$& 	$87$&	$70$&	$25$\\
$45$°&	$1.5$&	$30$&	$26.14$&	$1.26\cdot10^{17}$&	$68295$&	$28653$&	$8964$& 	$75$&	$71$&	$24$\\
$45$°&	$1.5$&	$0$&	$33.26$&	$1.59\cdot10^{17}$&	$373408$&	$2015$& 	$2703$& 	$71$&	$31$&	$58$\\
$45$°&	$1.5$&	$10$&	$33.27$&	$1.59\cdot10^{17}$&	$198845$&	$7061$& 	$7208$& 	$65$&	$30$&	$51$\\
$45$°&	$1.5$&	$30$&	$33.29$&	$1.58\cdot10^{17}$&	$61025$&	$12043$&	$13371$&	$59$&	$50$&	$53$\\
$45$°&	$1.5$&	$50$&	$33.27$&	$1.57\cdot10^{17}$&	$61977$&	$18379$&	$21015$&	$56$&	$42$&	$53$\\
$45$°&	$1.5$&	$10$&	$69.10$&	$3.19\cdot10^{17}$&	$111352$&	$1383$& 	$3467$& 	$48$&	$6.6$&	$72$\\
$45$°&	$1.5$&	$30$&	$69.11$&	$3.19\cdot10^{17}$&	$85448$&	$2918$& 	$8582$& 	$45$&	$7.3$&	$75$\\
$45$°&	$1.5$&	$50$&	$69.11$&	$3.18\cdot10^{17}$&	$62910$&	$2797$& 	$12989$&	$39$&	$10.5$&	$82$\\
$75$°&	$1.5$&	$5$&	$11.97$&	$6.25\cdot10^{16}$&	$1081599$&	$74844$&	$4050$& 	$53$&	$43$&	$5.1$\\
$75$°&	$1.5$&	$10$&	$11.98$&	$6.23\cdot10^{16}$&	$492645$&	$111115$&	$7851$& 	$50$&	$54$&	$6.6$\\
$75$°&	$1.5$&	$30$&	$11.98$&	$6.07\cdot10^{16}$&	$304997$&	$103127$&	$10115$&	$43$&	$86$&	$8.9$\\
$75$°&	$1.5$&	$10$&	$15.44$&	$7.85\cdot10^{16}$&	$497902$&	$86481$&	$3716$& 	$60$&	$51$&	$4.1$\\
$75$°&	$1.5$&	$30$&	$15.44$&	$7.72\cdot10^{16}$&	$255930$&	$103781$&	$6287$& 	$53$&	$81$&	$5.7$\\
$75$°&	$1.5$&	$10$&	$18.93$&	$9.46\cdot10^{16}$&	$461112$&	$56518$&	$1982$& 	$80$&	$50$&	$3.4$\\
$75$°&	$1.5$&	$30$&	$18.94$&	$9.36\cdot10^{16}$&	$168940$&	$76173$&	$4026$& 	$73$&	$91$&	$5.0$\\
$75$°&	$1.5$&	$10$&	$22.46$&	$1.11\cdot10^{17}$&	$422752$&	$34474$&	$3763$& 	$87$&	$53$&	$9.8$\\
$75$°&	$1.5$&	$30$&	$22.46$&	$1.10\cdot10^{17}$&	$126971$&	$54663$&	$6935$& 	$79$&	$72$&	$11.3$\\
$75$°&	$1.5$&	$10$&	$26.00$&	$1.27\cdot10^{17}$&	$403659$&	$21127$&	$5293$& 	$76$&	$30$&	$20$\\
$75$°&	$1.5$&	$30$&	$26.01$&	$1.26\cdot10^{17}$&	$101761$&	$36738$&	$11634$&	$71$&	$64$&	$24$\\
$75$°&	$1.5$&	$10$&	$33.10$&	$1.59\cdot10^{17}$&	$501497$&	$14368$&	$4959$& 	$62$&	$10.2$&	$26$\\
$75$°&	$1.5$&	$30$&	$33.11$&	$1.59\cdot10^{17}$&	$111518$&	$24493$&	$21310$&	$58$&	$41$&	$47$\\
$75$°&	$1.5$&	$50$&	$33.11$&	$1.58\cdot10^{17}$&	$84322$&	$24828$&	$22230$&	$56$&	$42$&	$47$\\
\end{longtable}
}
\end{document}